# Coherent manipulation of four-level double Λ-like atomic system by a train of ultra-short few-cycle-optical pulses

**Pawan Kumar and Amarendra K. Sarma***
Department of Physics, Indian Institute of Technology Guwahati, Guwahati-781039, Assam, India.
*Electronic address: aksarma@iitg.ernet.in

We have demonstrated that near complete coherence can be achieved in a four level double Λ-like systems using a train of ultra-short pulses. The effect of the Doppler broadening has been analyzed and a scheme has been proposed for establishing high and uniform coherence across different velocity groups in the atomic ensemble. We have also presented a novel scheme of excitation using chirped pulses and shown that in addition to generating coherence in the system it is possible to alter the translational states of the atoms.

## I. INTRODUCTION

Since the first experimental demonstration of coherent population trapping (CPT) in Λ-type system [1,2], numerous investigations have been made into coherence induced phenomena like electromagnetically induced transparency (EIT), absorption (EIA), etc. [3-6].Arguably, one of the primary reasons behind such an interest is the possibility of modifying the optical properties of the media through quantum interference. Extremely narrow absorption and emission spectral features [7] are associated with the phenomena and have been identified for use in frequency standards [8-9]. The refractive index of the medium shows steep normal variation near the dark resonance region and leads to large reductions in group velocity of light [10].The coherence accumulated in a system can give rise to effects like enhancement of refractive index with zero absorption [11-13], light amplification without population inversion [14], high-harmonic generation [15],etc. The possibility of using a train of pulses to create coherence was predicted by Kocharovskaya and Khanin in 1986 [16] and later demonstrated experimentally by making use of a comb of optical pulses produced from a mode-locked diode laser [17].With the recent progress in generation of well characterized short optical pulses it has become possible to accumulate the coherence in a robust and controllable fashion [18-19]. More recently authors have made theoretical investigations into the generation of coherence using a series of short pulses [20-21]. However the accumulation of coherence using a series of ultra-short few-cycle pulses is relatively less explored. Further, interaction of femtosecond optical pulses has been shown to give rise to appreciable optical dipole forces with possible applications in focusing, defocusing and steering of neutral atoms[22-23]. It has been demonstrated by several authors that the dipole force induced by laser light can lead to creation of an optical lens for the neutral atomic and molecular beam [24-26].Application of a series of tailored ultra-short pulses to simultaneously create coherence and manipulate the translational state of neutral atomic system through induced optical dipole forces is however much less explored.In this work we present a detailed analysis of coherence creation between the two lower states of a four level double Λ-like atomic system using a train of ultra-short few-cycle Gaussian pulses. The effect of Doppler broadening on coherence accumulation process has been analyzed. We also present a scheme to control the trajectory of atoms using the optical dipole force induced by the pulse train along with generation of coherence in the system. The article is organized as follows. In Sec. II we present the density matrix equations describing the interaction of the four-level system with a train of femtosecond pulses. Sec. III contains the simulated results of the interaction process and their analysis, followed by conclusions in Sec. IV.

## II. THE MODEL

The atomic system considered in this work for creation of coherence is the four-level system shown in Fig.1.These levels can easily be accessed in experiments by considering the well-studied $D_1$ transition hyperfine structure in alkali atoms like Cs, Rb, Na,etc. [27]. For example, state |1> and |2> could correspond to F=2 and F=3 hyperfine levels of the ground state $5S_{1/2}$ of Rb-85 respectively, whereas state |3> and |4> in that case would denote the F'=2 and F'=3 levels of $5P_{1/2}$. Here, we take a representative system and allow it to interact with a train of ultra-short, few-cycles femtosecond pulses of radiation. The physical situation corresponds to that of a beam of atoms in the gaseous state interacting with the co-propagating laser pulses [23]. A series of pulses in the time domain with well defined repetition period and phases gives rise to a comb in the frequency domain [28-29].Thus it is possible to excite a number of different transitions in the atomic system with the corresponding coherent coupling radiations using a single source of laser pulses.

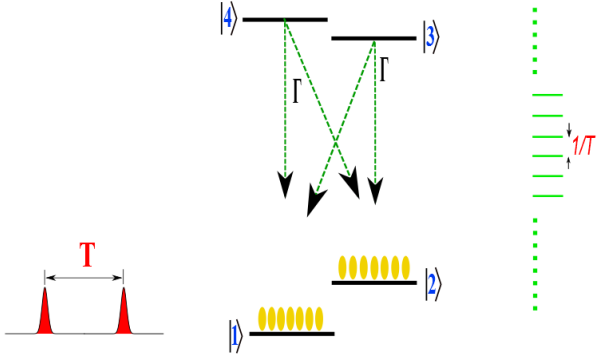

FIG. 1.Schematic diagram of the energy levels and the coupling radiation

The electric field envelope of the laser pulse-train interacting with the atomic system is described by $G(r,t) = \sum_{n=0}^{n=N-1} E_0 g(r, t - nT) e^{in\Delta\phi}$ where $g(r,t) = \exp\left(-\left(\left(\frac{r}{W_0}\right)^2 + \left(\frac{t}{\tau_0}\right)^2\right)\right)$ is the Gaussian envelope of a single pulse in the train with the peak amplitude $E_0$ and $\Delta\phi$ is the pulse to pulse phase shift. $W_0$ and $\tau_0$, respectively are the spatial and temporal widths of an individual pulse. $T$ is the pulse repetition period for the train. Within a pulse the electric field profile is given by $E(r,t) = E_0 g(r,t) Cos(\omega_0 t - \Phi(z))$, where $\omega_0$ is the central frequency of the laser pulse. The equation of motion for the density matrix elements describing the time evolution of the considered system is presented in Eq1.The well known rotating wave approximation breaks down in the case of interaction with a few cycle pulses and thus Eq.1 is derived without invoking the RWA [30].

$$\frac{d\rho_{11}}{dt} = (Y_{41}\rho_{44} + Y_{31}\rho_{33}) + i(\Omega_{13}\rho_{31} - \Omega_{31}\rho_{13}) + i(\Omega_{14}\rho_{41} - \Omega_{41}\rho_{14})$$

$$\frac{d\rho_{22}}{dt} = (Y_{42}\rho_{44} + Y_{32}\rho_{33}) + i(\Omega_{23}\rho_{32} - \Omega_{32}\rho_{23}) + i(\Omega_{24}\rho_{42} - \Omega_{42}\rho_{24})$$

$$\frac{d\rho_{33}}{dt} = -(\gamma_{31} + \gamma_{32})\rho_{33} + i(\Omega_{31}\rho_{13} - \Omega_{13}\rho_{31}) + i(\Omega_{32}\rho_{23} - \Omega_{23}\rho_{32})$$

$$\frac{d\rho_{44}}{dt} = -(\gamma_{41} + \gamma_{42})\rho_{44} + i(\Omega_{41}\rho_{14} - \Omega_{14}\rho_{41}) + i(\Omega_{42}\rho_{24} - \Omega_{24}\rho_{42})$$

$$\frac{d\rho_{12}}{dt} = i(\omega_{21}\rho_{12} + \Omega_{13}\rho_{32} + \Omega_{14}\rho_{42} - \Omega_{32}\rho_{13} - \Omega_{42}\rho_{14})$$

$$\frac{d\rho_{13}}{dt} = -\left(\frac{\gamma_{31}}{2} + \frac{\gamma_{32}}{2}\right)\rho_{13} + i(\omega_{31}\rho_{13} + \Omega_{13}(\rho_{33} - \rho_{11}) + \Omega_{14}\rho_{43} - \Omega_{23}\rho_{12})$$

$$\frac{d\rho_{14}}{dt} = -\left(\frac{\gamma_{41}}{2} + \frac{\gamma_{42}}{2}\right)\rho_{14} + i(\omega_{41}\rho_{14} + \Omega_{14}(\rho_{44} - \rho_{11}) + \Omega_{13}\rho_{34} - \Omega_{24}\rho_{12})$$

$$\frac{d\rho_{23}}{dt} = -\left(\frac{\gamma_{31}}{2} + \frac{\gamma_{32}}{2}\right)\rho_{23} + i(\omega_{32}\rho_{23} + \Omega_{23}(\rho_{33} - \rho_{22}) + \Omega_{24}\rho_{43} - \Omega_{13}\rho_{21})$$

$$\frac{d\rho_{24}}{dt} = -\left(\frac{\gamma_{41}}{2} + \frac{\gamma_{42}}{2}\right)\rho_{24} + i(\omega_{42}\rho_{24} + \Omega_{24}(\rho_{44} - \rho_{22}) + \Omega_{23}\rho_{34} - \Omega_{14}\rho_{21})$$

$$\frac{d\rho_{34}}{dt} = -\left(\frac{\gamma_{32}}{2} + \frac{\gamma_{31}}{2} + \frac{\gamma_{42}}{2} + \frac{\gamma_{41}}{2}\right)\rho_{34} + i(\omega_{43}\rho_{34} + \Omega_{31}\rho_{14} + \Omega_{32}\rho_{24} - \Omega_{14}\rho_{31} - \Omega_{24}\rho_{32} \quad (1)$$

In these equations $\gamma_{ij}$, $\omega_{ij} = \omega_i - \omega_j$ and $\Omega_{ij}$ denote the spontaneous decay rate, difference in energy levels and the Rabi frequency corresponding to the pulsed radiation respectively for the energy level pair |i> and |j>. We have assumed that the pair of transitions |1> to |2> and |3> to |4> are dipole forbidden and the rate of decoherence between the two ground states |1> and |2> is negligible and has no influence over the time-scales involved in the interaction with the pulse train. It is very instructive to express the transition frequencies in terms of the repetition frequency, $\nu = 1/T$ of the pulse train and hence we write $\omega_{ij} = 2\pi n_{ij}\nu$ where $n_{ij}$ is a dimensionless number. The Rabi frequencies for various electronic transitions are given by $\Omega_{ij} = \mu_{ij}E(r,t)/\hbar$ where $\mu_{ij}$ is the electric dipole moment for the transition |i> to |j>. We have the relationship $\Omega_{ji} = \Omega_{ij}^*$.

### III. RESULTS AND DISCUSSIONS

A multitude of features is observed in the dynamics of the double-Λ system as a result of the interaction with a train of ultra-short pulses. Depending on the frequencies of electronic transitions available in the atomic system and the relative position of the comb frequencies, the system can be driven to different final states. We show the time evolution of the atomic states in Fig.2 under four different conditions of excitation. These results are obtained by numerical integration of Eq.1 using a fourth order Runge-Kutta scheme. The values of parameters used in the simulation are: $\gamma_{41}=\gamma_{42}=\gamma_{31}=\gamma_{32}= \Gamma/2 =25\times10^6 s^{-1}$. This corresponds to a lifetime of 20 ns for the excited states. We also assume that $\Omega_{14}=\Omega_{13}=\Omega_{24}=\Omega_{23}=\Omega$ and that $\Omega_{ij}$'s are all real for simplicity in calculations. Each pulse in the train is characterized by a temporal width given by $\tau_0 = 10 fs$ and spatial width corresponding to $W_0 = 100 \mu m$. We choose a representative off axis point at $R = 50 \mu m$ for presenting these results. The peak Rabi frequency is taken to be $\frac{\sqrt{\pi}}{200} rad/fs$, which corresponds to a pulse area of $\frac{\pi}{20}$. The pulse-to-pulse phase-shift, $\Delta\phi$, is assumed to be zero in the simulation. In Fig.2.(a) we depict the time evolution of the atomic system for the case when the repetition frequency $\nu = 100$ MHz and the energy levels are given by $n_{41} = 3.75 \times 10^6, n_{21} = 30$ and $n_{43} = 3.6$. The system starts with equal populations in state |1> and |2>, $\rho_{11} = \rho_{22} = 0.5$ and zero initial coherence, $\rho_{12} = 0$. The repetition frequency is such that the condition for two-photon resonance is fulfilled and the system reaches to the maximum coherence state after interaction with about 150 pulses.

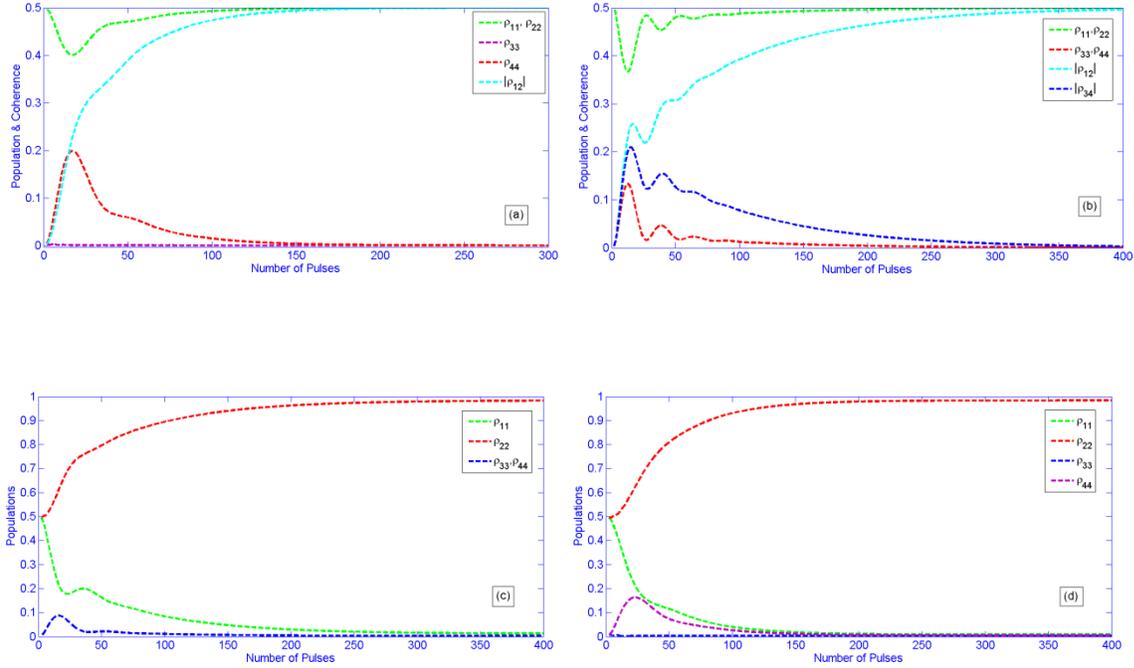

FIG 2. Evolution of populations and coherences under various conditions of excitation (a) ν=100 MHz (b) ν=120 MHz (c) ν=90 MHz and (d) ν=109 MHz

At this stage nearly all the atoms are in a superposition state of level |1> and |2> given by $\frac{|1>-|2>}{\sqrt{2}}$. One can also observe that population is preferentially pumped into the level |4> and $\rho_{33}$ remains close to zero during the entire interaction process. Thus it is possible to isolate the Λ-system formed by the levels |1>, |2> and |4> with very high selectivity even though the frequency spread of a single pulse considered in the train is a few tens of THz. Fig.2.(b) displays the case in which two photon resonance condition is satisfied with $\nu = 120$ MHz for both the pairs of energy levels ( |1> ,|2> ) and (|3>,|4>) with $n_{21} = 25$ and $n_{43} = 3$. We note that a considerable amount of coherence develops between the upper two states in this case in addition to coherence creation between the lower two states in the beginning of the interaction with the pulse train. It is also evident that the modulation in the populations and coherences is much more pronounced in this case as compared to Fig.2.(a). These modulations are a result of the Rabi oscillations of population between ground and excited states[20]. The scheme however is not very efficient with respect to coherence creation between the two lower states as it takes more than 300 pulses to achieve the full coherence. Fig.2.(c) and (d) present the results for $\nu = 90$ MHz and $\nu = 109$ MHz respectively. A repetition frequency of 90 MHz translates into $n_{43} = 4$ and $n_{21} = 100/3$. Thus levels |3> and |4> are in two photon resonance but the two lower states are not. We observe that population of both the upper states builds up in the beginning but the system is finally driven to a state with near complete (~99%) population accumulation in state |2>. This is a manifestation of efficient optical pumping. When both $n_{43}$ and $n_{21}$ are non-integers, as is the case with $\nu = 109$ MHz, optical pumping into the upper ground state |2> is highly efficient. The central frequency of the laser was tuned in resonance with $\omega_{41}$ in all of these simulations. The magnitude of Rabi Frequency determines the rate of coherence creation between the ground states of a Λ-type system with respect to the number of pulses required in the train. It has been

shown that a larger Rabi Frequency leads to complete coherence creation with relatively less number of pulses [20,21].For the Gaussian pulse profile as considered in this work, amplitude of the electric field and hence the associated Rabi frequency decreases exponentially as one moves away from the common axis of the atomic beam and pulse train. We display in Fig.3 the accumulation of coherence, $|\rho_{12}|$, at different radial locations within the atomic beam cross-section as a function of the number of femtosecond pulses in the train.

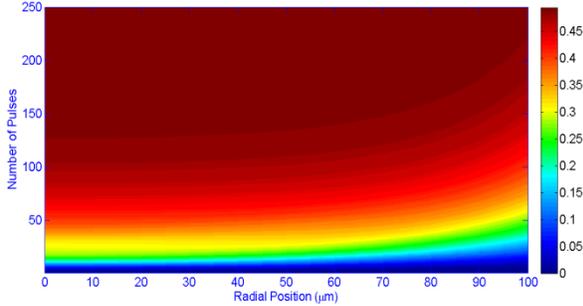

FIG. 3. Variation of coherence accumulated in a collimated atomic beam over the cross-sectional plane.

One can observe that despite the exponential decrease of the Rabi frequency, it is possible to obtain the maximum coherence ($|\rho_{12}| \approx 0.5$) over a substantial part of the atomic system with a limited number of pulses in the train. The other parameters taken for this simulation are the same as those used in Fig.2 (a).It is clear from Fig.2.(b) that the presence of an additional upper level can affect the rate of accumulation of coherence between the two ground states in a double Λ-like system. In particular, under specific conditions of excitation the number of pulses required for appreciable coherence creation can be significantly increased. Thus it is worthwhile to study the effect of the additional upper level on the coherence accumulation process in a systematic manner. We assume that the Λ-type system consisting of levels |1>,|4> and |2> is simultaneously in one photon and two photon resonance. The influence that level |3> has on $\rho_{12}$ in this configuration as it comes in and goes out of resonance is depicted in Fig.4 below.

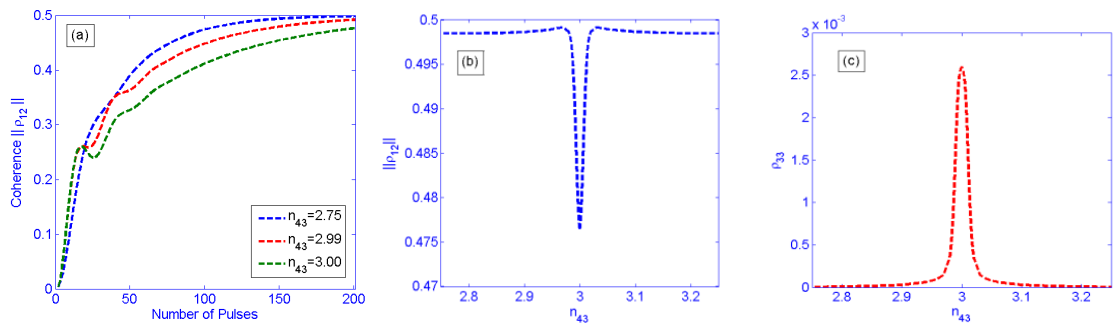

FIG.4.Dependence of coherence $|\rho_{12}|$ on n43 (a) pulse-to-pulse accumulation (b) final coherence and (c) final population $\rho_{33}$ after interaction with 200 pulses.

Fig.4.(a) displays the pulse-to-pulse build of coherence for three different relative values of $n_{43}$.Other atomic and pulse-train parameters have been assumed to be the same as in Fig.2.(a) for easy comparison and investigation. Here $n_{43} = 3.00$ represents the case when the level |3> is in exact resonance where as $n_{43} = 2.99$ describes a situation when the level is off resonance by 1 MHz. We note that the magnitude

of Rabi Oscillations increase as state |3> comes in resonance and although the rate of coherence accumulation is faster in the beginning of the train the convergence of $|\rho_{12}| \to 0.5$ is quite slow as compared to the far-off resonance case of $n_{43} = 2.75$. As we have shown earlier, in a real atomic system these situations could be realized by varying the repetition rate of the pulses in the train. Fig.4.(b) and (c) show the coherence, $\rho_{12}$, and population, $\rho_{33}$ respectively after interaction with 200 pulses in the train. We find a dip in coherence and a spike in the population of level |3> centered at the exact resonance position. The full width at half maxima (FWHM) of these Lorentzian structures is found to be of the order of about 2 MHz. It is important to note here that the natural line widths of the excited states considered in this work are greater than the FWHM found above. In conclusion, it is possible to judiciously choose the repetition frequency of the pulses so as to drive the system to near full coherence with relatively less number of pulses in the train. Up to this point the excitation pulses considered for study were taken such that the repetition period, $T$, in the train is less than the lifetimes of the excited states. Thus before the system is able to relax completely to the ground states it is subjected to the next pulse in the train. We now turn our focus to the regime where $T \geq T_{Decay}$. Here $T_{Decay}$ refers to the decay life-time of the excited states. We define an average Rabi Frequency, $\Omega_{avg}$, as $\Omega_{avg} = \{\int \Omega(t)\,dt\}/T$ where the integral is taken over a single pulse of the train. For the simulations we keep $\Omega_{avg}$ unchanged and equal to the case presented in Fig.2.(a) and vary the repetition period $T$ beyond the $T_{Decay}$. The results obtained from such an interaction are shown in Fig.5.

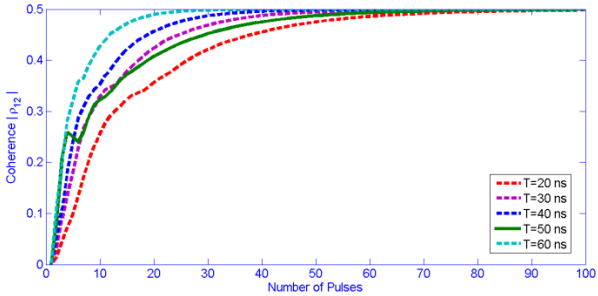

FIG.5. Build-up of coherence $|\rho_{12}|$ in the system with pulse repetition period $T \geq T_{Decay}$.

One may notice that pulse trains with increasing repetition period remain efficient in driving the system to the maximum coherence point. We observe that with increase in the repetition period of the pulses the number of pulses required to create the coherence decreases accordingly as the average Rabi Frequency has been kept constant. For the pulse train with $T = 60\ ns$, about 25 pulses are sufficient to generate the full coherence between |1> and |2>. A deviation is observed for the excitation with repetition period of $50\ ns$ as it takes considerably larger number of pulses to arrive to the optimum coherence point. We attribute this feature to the fact that for the corresponding repetition frequency, level |3> is in resonance with $n_{43} = 18$. This leads to increased oscillation of the coherence and slower convergence rate. Motivated by the efficiency of pulses with large repetition periods in coherence accumulation process we now explore the possibility of using a few high-energy pulses to create coherence in the atomic system. Femtosecond pulses with pulse areas as large as a few $\pi$'s has been considered by authors in the context of coherent population transfer in a multi-level atomic system [31]. Here we consider the treatment of the double Λ-type system with pulses for which the peak Rabi Frequency is given by $\sqrt{\pi}/20\ rad/fs$. Again the atomic and other pulse parameters are assumed to be the same as in Fig2.(a) for uniformity. Fig.6 displays the accumulated coherence $\rho_{12}$ versus the number of pulses for five different repetition periods in the regime with $T > T_{Decay}$. We note that the proposed scheme is able to generate high coherence (~98 % and above) with less than 15 pulses in the train. Another point to be noted is that unlike the convergence scenario in the case of Fig.4.(a) where the rate of accumulation becomes very low as $|\rho_{12}|$ approaches 0.5 , here the transition to full coherence remains fast enough during the later stages as well.

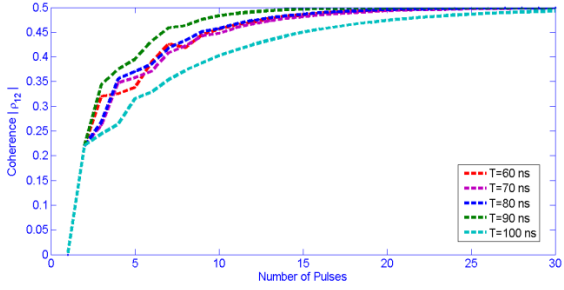

FIG. 6. Coherence generation using pulse train made up of a few pulses.

A real atomic ensemble at room temperature consists of groups of atoms moving with high velocities. Due to this motion the transition frequencies in the atom are shifted by an amount $\vec{k}.\vec{V}$ in the laboratory frame. Here $\vec{k}$ is the wave vector of the incident radiation and $\vec{V}$ denotes the velocity of the atom. Under the conditions of thermal equilibrium the spread of the velocity distribution, which is Gaussian in nature, can typically be as large as a few hundreds of MHz. This can lead to significant modifications in the response of the system. In this section we present the effects of the Doppler shift on the process of coherence creation in the double Λ-like atomic systems. We consider a repetition frequency of $\nu = 100$ MHz for the excitation pulses with peak Rabi Frequency of $\frac{\sqrt{\pi}}{50}\ rad/fs$. The central frequency of the laser pulses is taken to be $\omega_0 = (2\pi)3.75 \times 10^6 \nu$. This corresponds to a detuning of 1.25 MHz for each $1\ m/s$ speed of the atoms. As we mentioned earlier, the geometry of interaction is assumed such that the atomic beam co-propagates with the pulses and thus the longitudinal velocity of the atoms is responsible for the Doppler shift of the transition frequencies.

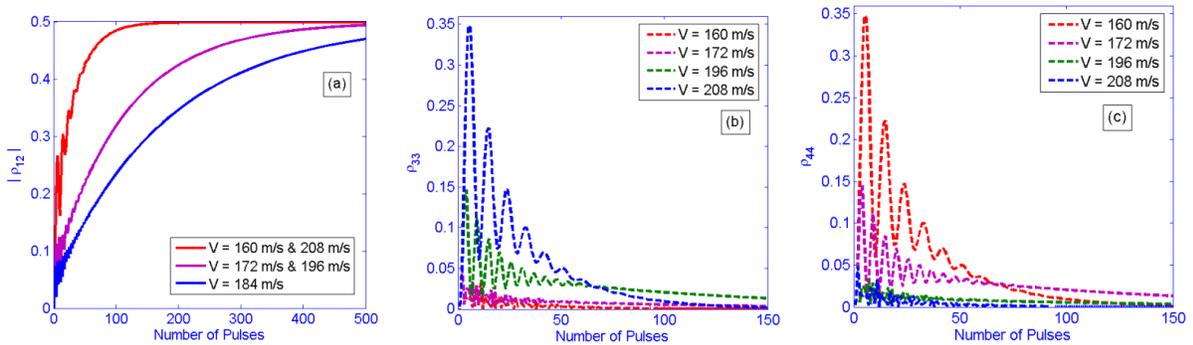

FIG.7. Effect of Doppler shift on (a) coherence creation and evolution of population (b) in state |3> and (c) in state |4>.

In Fig.7 (a) we show the coherence build up for five different longitudinal velocity groups in the ensemble. It is remarkable that the accumulation process is much faster for the groups with $V = 160\ m/s$ and $208\ m/s$ than the group with $V = 184\ m/s$. This is because whenever the system is in one photon resonance in addition to two photon resonance condition the coherence accumulation process is very fast and it takes relatively less number of pulses to drive the system to complete coherence. As is evident from Fig.7 (b) and Fig.7 (c), state |3> and state |4> alone are in one photon resonance with the two lower states for the velocity group $V = 208\ m/s$ and $V = 160\ m/s$ respectively. We note that the population in either of the excited states remains very low (~0.04) for the group with $V = 184\ m/s$. This group is far detuned from both the excited states with respect to one photon resonance condition and thus the coherence creation is quite slow. Fig.8 displays the result of interaction of the femtosecond pulses under the conditions considered in Fig.7 with a broadrange of velocity groups in the atomic ensemble as would be the case for a real atomic system.

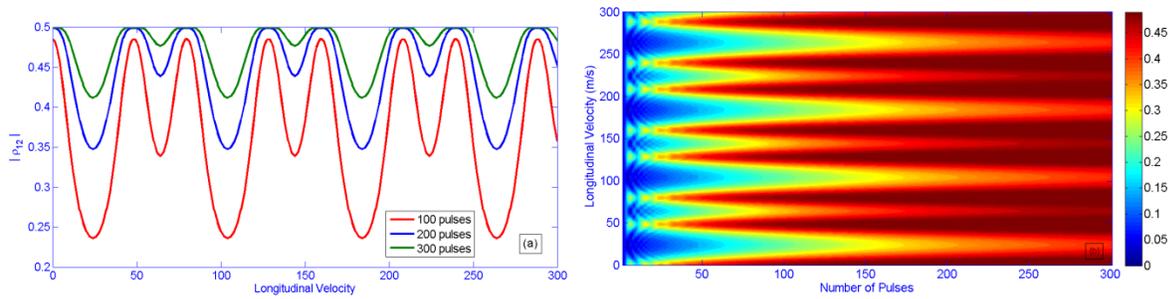

FIG. 8.(a) Variation in coherence induced for different velocity groups (b) Contour plot of variation in $|\rho_{12}|$ due to Doppler effect.

We can observe from Fig.8 that the coherence developed in the atomic system for a given number of excitation pulses varies periodically with the longitudinal velocity of the atoms. The coherence repeats itself after every $80\ m/s$. The behavior stems from the condition that the corresponding Doppler shift and the repetition frequency of the pulse train are equal. Hence in essence the conditions of excitation repeat after every $80\ m/s$ velocity because the separation between two successive teeth of the frequency comb is 100 MHz. We also note that within a period there are two maxima and minima of coherence. As we have explained through Fig.7, the two maxima are a result of resonance condition fulfillment involving each of the two excited states. On the other hand we find that when both the upper quantum states are equally and far detuned from the teeth of the frequency comb, the coherence accumulated in the system attains minima. We have seen that Doppler shift of the energy levels can result into significant departures from near complete coherence for specific velocity groups in the atomic ensemble. Motivated by the results presented in Fig. 6, where few high energy pulses with large repetition periods were used to create the coherence in the system we now present the results obtained by considering the effect of Doppler broadening.

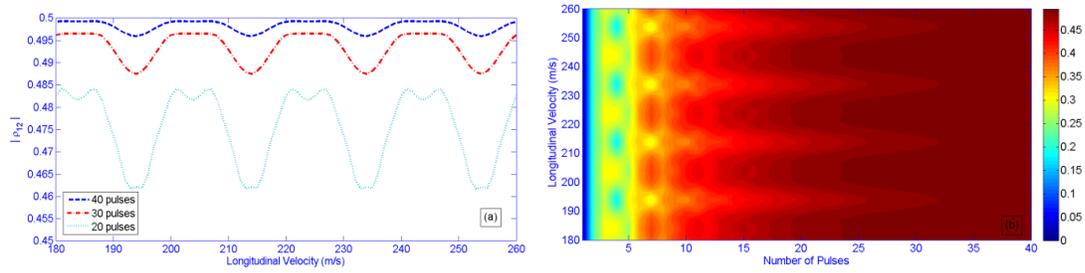

FIG. 9.Accumulation of $|\rho_{12}|$ for different velocity groups in the ensemble with the exciting train made up of a few pulses of large pulse areas.

Fig 9 (a) and (b) show the coherence $|\rho_{12}|$ accumulated by a pulse train with the repetition period of $40\ ns$. Other parameters were kept the same as in Fig.6 for this simulation. We find that with just 40 pulses it is possible to accumulate $|\rho_{12}|$ to values more than 99 % in all the velocity groups. It is remarkable that the dips in coherence as it varies periodically with the longitudinal velocity are very small. The suggested scheme may prove important for practical applications where a very high and uniform coherence is required in the entire system.

As an atom interacts with a pulse, the off-diagonal coherence terms of the density matrix are excited. Due to the non-zero gradient of the amplitude of the electric field for a focused Gaussian pulse the neutral atoms can experience a net dipole force. Studies on optical dipole force with chirped pulses have been carried out by authors where they have demonstrated pulse-induced forces on neutral atoms [23]. Here we now present an application of the scheme in the context of coherence accumulation in double Λ-like systems. We consider quadratic chirping of frequency of the individual pulses such that the instantaneous frequency is given by:$\omega(t) = \omega_0 + \chi t^2$, where $\omega_0$ is the central frequency and $\chi$ is the chirp rate. The time dependent optical dipole force on atoms interacting with the high energy ultra-short laser pulses is given by the expression

$$\vec{F}_r = \{\mu_{14}(\rho_{14} + \rho_{41}) + \mu_{24}(\rho_{24} + \rho_{42}) + \mu_{13}(\rho_{13} + \rho_{31}) + \mu_{23}(\rho_{23} + \rho_{32})\}\{\sum_{n=0}^{n=N-1}[\nabla E_0 g(r, t - nT)] Cos[(\omega(t - nT))(t - nT) - \Phi(z)]\} \quad (2)$$

where symbols have their usual meaning. The gradient involved gives force in the radial direction due to the circular symmetry of geometry of interaction between co-propagating atomic beam and the laser pulses.

We consider both negative ($\chi < 0$) and positive ($\chi > 0$) quadratic chirping of the pulses and investigate into their effects on the translational motion of the atoms in the transverse plane. Fig. 10 (a) shows the coherence in the system for a pulse train made up of negatively chirped pulses.

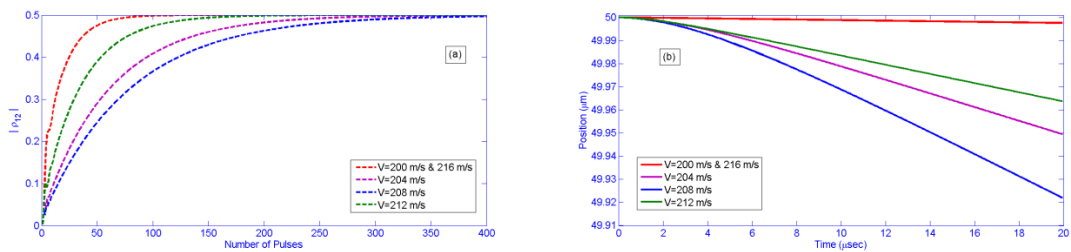

FIG. 10.(a) Evolution of $|\rho_{12}|$ with negatively chirped pulses (b) Position of atoms on the transverse plane.

We have chosen to display five different velocity groups of the ensemble that span a period excited by pulse train as described in Fig. 8. A remarkable feature of the interaction is that oscillations in $|\rho_{12}|$ during the accumulation process are significantly suppressed. This may be due to the fact that the central frequency of the pulse is no longer fixed and is swept in the pulse duration. The result was obtained by introducing a small quadratic chirp rate of $\chi = -10^{-3} fs^{-3}$. The atoms experience a converging force as a result of the interaction with the negatively chirped pulses and their translational states are altered as is clear from Fig. 10 (b). The repetition period of the tailored pulses was taken to be 50 ns and the Rabi Frequency was taken to be $\frac{\sqrt{\pi}}{4}$ $rad/fs$. We note that high energy pulses lead to appreciable deflections of the atomic trajectories. The atomic mass was taken to be $m = 85\ a.m.u$ for plotting the atomic trajectories. Another point to be noted is that amount of deflection depends on the velocity of the atoms. The effect stems from the fact that under a given condition of excitation the velocity group that reaches to the maximum coherence point with relatively less number of pulses experiences less deflection. This is because it stops interacting with light once full coherence is achieved in the system. For the velocity group with $V = 208\ m/s$ the rate of coherence accumulation is the slowest and thus it experiences the maximum deflection. On the other hand for groups with $V = 200\ m/s$ and $216\ m/s$ the amount of deflection is much smaller. We find that the pattern remains periodic even for the case of interaction with chirped pulses. In Fig.11 (b) we display the translational motion induced by the positively chirped pulses with $\chi = 10^{-3} fs^{-3}$. Other parameters were kept the same as in Fig. 10.

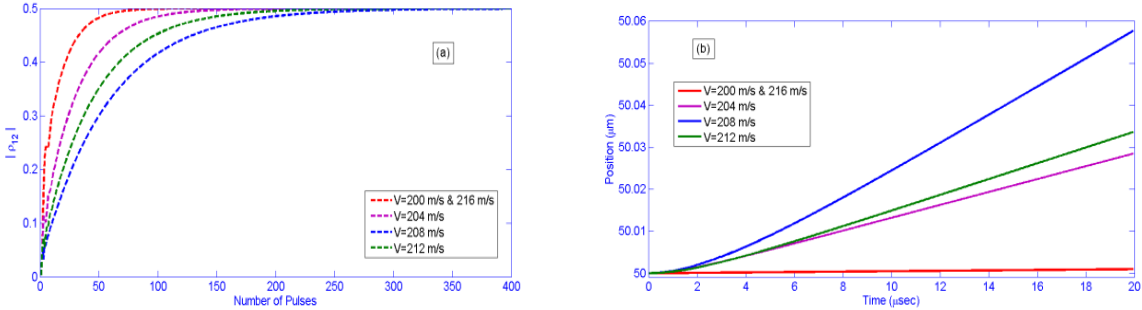

FIG. 11.(a) Evolution of $|\rho_{12}|$ with positively chirped pulses (b) Diverging nature of the forces on the atoms alters their translational state and position on the transverse plane.

The nature of the optical dipole force becomes opposite and leads to defocusing action. The dependence on the atomic velocity is similar to the case with negative chirping. The scheme presented here can be used for velocity selective deflections along with coherence creation. By varying the pulse repetition period it should be possible to select a few specific velocity groups for maximum deflections. The simulations for optical dipole force were carried out for an off-axis point located at a radial position of $50\ \mu m$.

## IV. CONCLUSIONS

In conclusion we have demonstrated that near complete coherence can be achieved in a four level double $\Lambda$-like systems using a train of ultra-short pulses. The conditions of excitation for efficient coherence accumulation are identified by investigating into the effect of the additional upper level. The effect of the Doppler shift has been analyzed and a scheme has been proposed for establishing high and uniform coherence across different velocity groups in the atomic ensemble. We have also presented a novel scheme of excitation using chirped pulses and shown that in addition to generating coherence in the system it is possible to alter the translational states of the atoms.

## References

[1] E. Arimondo, Progr. Opt, **35**, 257 (1996).
[2] G. Alzetta *et al*., NuovoCimento B **36**, 5(1976).
[3] K. J. Boller, A. Imamoglu, and S. E. Harris,Phys. Rev. Lett.,**66**, 2593 (1991).
[4] A.M. Akulshin, S. Barreiro, and A. Lezama,Phys. Rev. A,**57**, 2996 (1998).
[5] S. E. Harris,Phys. Today **50**, 36 (1997).
[6] A. Lezama *et al.*,Phys. Rev. A,**59**,4732 (1999).
[7] S. Brandt, A. Nagel, R. Wynands, and D.Meschede, Phys. Rev. A, **56**, R1063 (1997).
[8] J. Vanier,Appl. Phys. B **81**, 421 (2005).
[9] T. Bandi, C. Affolderbach, and G. Mileti, J.Appl. Phys**111**, 124906 (2012).
[10] O. Schmidt, R. Wynands, Z. Hussein, andD. Meschede, Phys. Rev. A, **53**, R27(1996).
[11]M. O. Scully, Phys. Rev. Lett. **67**, 1855, (1991).
[12]M. Xiao *et al*., Phys. Rev. Lett. **74**, 666 (1995).
[13]M. O. Scully and M. Fleischhauer,Phys. Rev.Lett. **69**, 1360 (1992).
[14] S. E. Harris,Phys. Rev. Lett. **62**, 1033 (1989).
[15] J. B. Watson, A. Sanpera, X. Chen and K. Burnett, Phys. Rev. A **53**, R1962 (1996).
[16] O. A. Kocharovskaya and Y. I. Khanin, Sov. Phys. JETP **63**, 945 (1986).
[17] Vladimir A. Sautenkov *et al*,Phys. Rev. A **71**, 063804 (2005).
[18] D. Aumiler, Phys. Rev. A **82**, 055402 (2010).
[19] Matthew C. Stowe, AviPe'er, and Jun Ye, Phys.Rev. Lett. **100**,203001 (2008).
[20] A. A. Soares and Luís E. E. de Araujo, Phys. Rev. A **76**,043818 (2007).
[21] Marco P. Moreno and Sandra S. Vianna,J.Opt.Soc.Am.B,Vol.**28**, No.5,1124 (2011).
[22] P. Kumar and A. K. Sarma, Phys. Rev. A **84**,043402 (2011).
[23] P. Kumar and A. K. Sarma, Phys. Rev. A **86**,053414 (2012).
[24] O. Steuernagel, Phys. Rev. A **79**, 013421 (2009).
[25] G. M. Gallatin and P. L. Gould, J. Opt. Soc. Am.B**8**, 502 (1991).
[26] B.S.Zhao *et.al.*,Phys. Rev. Lett. **85**, 2705 (2000).
[27] Reference data for alkali atoms compiled by D. A. Steck.
[28] J.N. Eckstein, A. I. Ferguson, and T.W. Hänsch, Phys. Rev. Lett. **40**, 847 (1978).
[29] L. Xu et al., Opt. Lett. **21**, 2008 (1996).
[30] T. Nakajima and S. Watanabe, Phys. Rev. Lett. **96**, 213001(2006).
[31]P. Kumar and A. K. Sarma, Phys. Rev. A **88**, 033823 (2013).